\documentstyle[prb,aps,preprint]{revtex}
\begin{document}
\draft
\title{Structural and Magnetic Properties of Trigonal Iron}
\author{S. Fox and H.J.F. Jansen}
\address{Department of Physics, Oregon State University, Corvallis, OR
97331, USA}
\date{\today}
\maketitle

\begin{abstract}
First principles calculations of the electronic structure of trigonal iron
were performed using density function
theory.  The results are used to
predict lattice spacings, magnetic moments and elastic properties;
these
are in good agreement with experiment for both the bcc and fcc
structures.
We find however, that in extracting these quantities great care must be
taken in interpreting numerical fits to the calculated total energies.  In
addition, the results for bulk iron give insight into the properties
of thin iron films.
Thin films grown on substrates with mismatched lattice constants often
have non-cubic symmetry.  If they are thicker than a few monolayers
their electronic structure is similar to a bulk material with an
appropriately distorted geometry, as in our trigonal calculations.
We recast our bulk results in terms of an iron film grown on the
(111) surface of an fcc substrate, and find the predicted strain
energies and moments accurately reflect
the trends for iron growth on a variety of substrates.
\end{abstract}


\narrowtext

\section{Introduction}

    Thin transition metal films can be grown on a variety of
substrates.  The position of the atoms in the first film layer is
largely determined by the surface geometry of the substrate.
As more layers are added the lattice geometry of the film is
determined by the forces at the substrate interface and by the
inter-atomic forces within the film.  As the film gets even thicker
its geometry will eventually become that of the stable bulk material.
Films with just a few layers give us the opportunity to study
transition metals in geometries other than that of the bulk
material, but with the additional complication of a interface with the
substrate.
Electronic structure calculations that include the effects of the
substrate interface have revealed that other than the imposition
of a non-bulk lattice geometry throughout the film, only the
interface layer of the metal film is strongly affected by the
substrate.\cite{FuFreeman}  As a result bulk electronic structure
calculations which ignore the details of the interface interactions
but which use lattice constants characteristic of the substrate
should give a good picture of the thin film electronic structure.

     Thin films of iron can be grown on substrates with a variety
of lattice constants, so a range of calculations, corresponding to
several different distortions of the bulk structure, are necessary
to characterize these films. A distortion of a cubic unit cell along
the (111) axis can be examined using a trigonal basis.  The trigonal
basis, characterized by three equal length vectors originating at the
origin and the equal angle which separates each of them, contains the
bcc, fcc, and simple cubic structures as special cases.  By varying the
characteristic angle one can obtain these special geometries and the
distortion along the (111) axis which connects them.  Electronic
structure calculations of distorted structures near the bcc and fcc
geometries will allow us to extrapolate elastic constants for these
bulk geometries. More importantly in this context, these trigonal
structures are exactly those which match the lattice provided by the
(111) face of an fcc substrate.  So an electronic structure
calculation of bulk iron with a trigonal geometry gives us
insight into iron films grown on these substrates.  Similarly, a
tetragonal unit cell provides the correct distortion to match the
(100) face of a cubic substrate.\cite{PengJansen}

     In this brief report we will give the results of electronic
structure calculations for ferromagnetic bulk iron over a range of
trigonal geometries.  The full-potential linearized augmented plane
wave method (FLAPW) was used with Janak's
parametrization of von Barth and Hedin's local density
approximation to the exchange and correlation potential.
\cite{JansenFreeman,vonBarthHedin}  Some refinement of
results from similar
previous calculations for a tetragonal geometry will also be presented.
\cite{PengJansen}

\section{Results for Bulk Materials}

	Figure \ref{figure1} shows the range of lattice geometries
over which the
energy calculations were performed. For this one atom per unit cell
trigonal structure the geometry is parameterized in
terms of the c/a ratio and the volume per
unit cell.  The
natural log of the c/a ratio is used since it provides a more natural
scaling.  The volume is scaled by the experimental bcc volume
(78.83 a.u.).  The three special geometries,
fcc, simple cubic, and bcc, correspond to ln(c/a) values of 0.69, 0.0
and -0.69 respectively.  Geometries at which electronic structure
calculations were performed are indicated by a dot.  Using the
total energy result at each of these points we can numerically
construct constant energy contours as shown in figure \ref{figure1}.

	The most prominent features in this contour plot are the two
local energy minimas corresponding to an fcc and a bcc structure.  As
has been observed in previous calculations
the global minima is mispredicted; the fcc structure lies
about 4 mRyd low than the bcc structure.  At lower temperatures iron
is actually a bcc crystal so we would expect a bcc structure to
appear as the global minima for these calculations.  This error is
widely attributed to the inherent inaccuracy of the specific form of
the local density approximation we have used.\cite{JansenPeng}
One can travel between these two locally stable structures by varying
the c/a ratio, which corresponds to distorting
either cubic structure along the (111) axis.
The energy barrier between bcc and fcc structures for this distortion
is approximately 70 mRyd/atom, significantly larger than the
10 mRyd/atom found for the tetragonal distortion which connects
the bcc and fcc structures. \cite{PengJansen}

     The data near the two minima can be used to obtain lattice
constants, volumes and elastic properties of these two high symmetry
structures.  Ideally one would investigate regions extremely
close to the energy minima where the energy varies quadratically with
small distortions in the geometry.  In this case a simple quadratic
fit would perfectly model the behavior of the electronic energy around
the minima allowing the various crystal properties to be extracted with
great reliability.  Unfortunately,  because of the limited accuracy
in the calculated energies, we need to look at structures further from
the minima to ensure the energy differences among our structures are
larger than the background numerical noise.  As a result, we can no
longer expect a simple quadratic fit to perfectly model variation
in total energy among the structures near the minima.

	To account for this inherent inaccuracy the data were fit by
a number of methods.  Two dimensional fits of energy versus volume
along the fcc $(\ln (c/a)=0.69)$ and bcc $(\ln (c/a)=-0.69)$ lines were
performed with a simple quadratic model as well as with the
Murnaghan and Birch-Murnaghan models.\cite{BirchMurnaghan}
Three-dimensional quadratic fits were made to points near the minima
in the c/a versus V plane.  The particular points to which each model
was fitted were varied and the sensitivity of the final results to
these variations was noted.  In all cases
`good' (in terms of Chi-squared) fits were obtained. The variation in
results between the different fits was used as an
estimate of the uncertainty introduced by the use of imperfect fitting
models. The lattice
constants and volumes obtained were consistent for the various
fitting techniques.  The elastic constants showed some variation,
especially for the fcc minima.  The elastic constants obtained from
the three-dimensional quadratic fit were especially sensitive to the
selection of fitted points.  Tables \ref{table1} and \ref{fcctable}
show some sample results from
a variety of fits done near the bcc and fcc minima.  Considered
carefully, the fitting errors are no greater than those known to be
inherent in the local density approximation: 2-3\% in the lattice
constants and 10-20\% in the elastic constants.

	Combining these fit results and similar results from a
refitting of previous calculations on tetragonally distorted iron,
we obtained the predicted bulk ferromagnetic iron
properties shown in table \ref{table2}.\cite{PengJansen}  Along with
our results, table \ref{table2} gives
experimental numbers for these properties along with the results from
other density functional
calculations.\cite{GaoJohnstonMurrell,WangKleinKrakauer}
The numbers for the bcc minima are in good
agreement with the other theoretical predictions (also done under
the LDA)
consistently underestimating the lattice constants by a few percent,
and
over-estimating the elastic properties by a few tens of percent.  For
the fcc data we present theoretical numbers from the so called
`low-spin' ferromagnetic state and experimental results from
gamma-iron, the high temperature fcc phase.   The discrepancy between
theory and experiment is larger in this comparison, which is expected
since the theoretical calculations do not take temperature
fully into account.  In both the fcc and bcc case the large values for
the elastic constants can be largely attributed to the underestimate
of the lattice constant.  In general one finds in LDA calculations
that the smaller the equilibrium volume, compared to the experimental
value, the larger the bulk modulus.  If we were to force the unit cell
volume to be closer to the experimental value we could obtain elastic
properties closer to those experimentally measured.

\section{Relation to Thin Films}

	As was mentioned above, these bulk results can be related to
thin iron films by recasting the data in terms of iron grown
in perfect registry with the (111) surface of an fcc substrate.
In figure \ref{figure2} we show a contour plot of the same data as
figure \ref{figure1}, but
now it has been parameterized in terms of a film geometry. The
in-plane lattice spacing in the film will be fixed by the substrate
since we assume the film is grown pseudomorphically.  The interlayer
spacing will take on whatever value  minimizes the energy for the
given in-plane spacing.  If we know the lattice constant of our
substrate we can immediately extrapolate from figure \ref{figure2} the
interlayer
spacing that minimizes the energy.  Following this procedure for a
range of substrate lattice constants we construct figure
\ref{figure3} which
shows the energy of the film with optimal interlayer spacing as a
function of the substrate lattice constant.  The left and right
minima in this figure correspond to unstrained fcc, and unstrained
bcc growth
respectively.  The energy shift near these minima can be interpreted
as the strain energy per layer associated with strained, pseudomorphic
growth.

	As shown in figure \ref{figure4} we can, following a similar
procedure, express the results of
our bulk magnetic moment calculations as a function of
substrate lattice constant, again with optimal interlayer spacing.
In this figure we can see that unstrained fcc films have small
moments, but increasing the substrate lattice constant causes the
moment to rise.  In the region of bcc growth the moment is not
particularly sensitive to changes in the substrate.  Keep in mind
that these results ignore the interface effects which play a key
role in the magnetic properties of such films.

	With this caveat in mind we can relate these figures to the
observered properties of thin iron films for a variety of substrates.
Copper, with an fcc lattice constant of 6.82 a.u. should provide a
reasonable match for pseudomorphic growth of iron films.
Figures \ref{figure2}, \ref{figure3}, and \ref{figure4} predict iron
grown on copper (111) is a strained fcc
structure with an interlayer spacing of around $3.7 \pm 0.2$ a.u.
and a
moment of $1.0\pm 0.25$.  Recent studies showed strained fcc growth
up to five layers
with interlayer spacings of 3.9 a.u. and moments of $1\mu B$/atom
consistent with our result.\cite{TianJonaMarcus,RauSchneiderJamison}
Less well matched substrates have been
studied as well.  Using Pd , which has a lattice constant of
7.36 a.u., as a substrate should create a strain energy per layer
nearly twice
that of copper according to figure \ref{figure3}.  Not surprisingly
it was found that less than two layers of iron
could be grown pseudomorphically on Pd before islands of bcc iron
started to form.\cite{BegleyTianJonaMarcus}  Substrates which would
provide even larger strains
than Pd, namely Ag and Al do not appear to give clean
pseudomorphic growth as expected from figure \ref{figure3}.
\cite{BegleyTianJonaMarcus}

\section{Conclusions}

	In this paper we have examined a range of trigonal iron
structures using density functional theory.  We find energy minima
corresponding to the expected bcc and fcc geometries.  While carefully
considering the inherent inaccuracy in the fitting process we can
extract information about the properties of these bulk states which
are in good agreement with experimental results.  We also recast the
bulk results to reflect the geometries associated with the
pseudomorphic growth of thin iron films.  Although this approach
ignores the interface and surface effects it still gives meaningful
insight into the growth of iron films on a range of substrates.
The predicted film properties are consistent with the
growth observed for a variety of substrates.

\acknowledgments
This work was made possible by the Office of Naval Research under
grant N00014-9410326.

\begin{figure}
\caption{Contours of constant energy (mRyd/atom) of ferromagnetic
trigonal iron as a function of volume and c/a ratio. Dots
indicate geometries at which electronic structure calculations were
performed. All energies are offset by 2541 mRyd.}
\label{figure1}
\end{figure}

\begin{figure}
\caption{Contours of constant energy (mRyd/atom) of a ferromagnetic
iron film as a function of fcc substrate lattice constant and film
interlayer spacing. Dots indicate geometries at which electronic
structure calculations were performed. Contours far from data points
are unreliable.  All energies are offset by
2541 mRyd.}
\label{figure2}
\end{figure}

\begin{figure}
\caption{Energy (mRyd/atom) of a thin iron film with optimal interlayer
spacing as a function of the fcc
substrate lattice constant. All energies are offset by 2541 mRyd. A
line is provided to guide the eye.}
\label{figure3}
\end{figure}

\begin{figure}
\caption{Magnetic moment ($\mu B$/atom) of a thin iron film with
optimal interlayer spacing as a function of the fcc
substrate lattice constant.}
\label{figure4}
\end{figure}

\mediumtext

\begin{table}
\caption{Properties of bcc iron as predicted by a variety of fitting
models.\label{table1}}
\begin{tabular}{lllll}
c/a &	Volume (a.u.) &	Energy (mRyd.) & Bulk Modulus (Mbar) & Fitting
Method\\
\tableline
0.5 & 	73.4	&   -0.103 & 2.25 & parabola to 4 points\\
0.5 &	71.9	&   -0.102 & 1.77 & Murnaghan near minima\\
0.5 &	73.3	&   -0.103 & 2.30 & Birch-Murnaghan near
minima\\
0.5 &	71.1	&   -0.103 & 2.19 & Murnaghan over full
range of V\\
0.5 &	71.2	&   -0.104 & 2.33 & Birch-Murnaghan over
full range of V\\
0.5084 & 73.2	&   -0.101 & 2.11 & 3-d quadratic with
minimum error\\
0.5092 & 73.1   &   -0.101 & 2.14 & 3-d quadratic with
average error\\
0.5101 & 73.1   &   -0.102 & 2.22 & 3-d quadratic with
average error\\
\end{tabular}
\end{table}

\begin{table}

\caption{Properties of fcc iron as predicted by a variety of fitting
models.\label{fcctable}}
\begin{tabular}{lllll}
c/a &	Volume (a.u.) &	Energy (mRyd.) & Bulk Modulus (Mbar) & Fitting
Method\\
\tableline
2.0 & 	65.8	&   -0.105 & 3.19 & parabola to 4 points\\
2.0 &	65.9	&   -0.109 & 3.09 & Murnaghan near minima\\
2.0 &	65.7	&   -0.105 & 3.16 & Birch-Murnaghan near
minima\\
2.0 &	66.0	&   -0.103 & 2.20 & Murnaghan over full
range of V\\
2.0 &	65.6	&   -0.104 & 2.57 & Birch-Murnaghan over
full range of V\\
2.023 & 67.2	&   -0.107 & 3.66 & 3-d quadratic with
minimum error\\
2.006 & 66.9   &   -0.111 & 3.74 & 3-d quadratic with
average error\\
2.014 & 66.9   &   -0.103 & 2.30 & 3-d quadratic with
average error\\
\end{tabular}
\end{table}

\begin{table}
\caption{Properties of bcc and fcc iron.  Results from this study and
other density functional calculations, as well as experimental results
for $\alpha$-iron and $\gamma$-iron. All elastic properties are in
Mbar.
\label{table2}}
\begin{tabular}{lllllll}
&lattice constant (a.u.)&moment ($\mu$B)&Bulk Modulus&$c_{11}
$&$c_{12}$ & $c_{44}$\\
\hline
bcc \\
theory &$5.24\pm 0.1$ &$2.0\pm 0.05$ &$2.2\pm 0.15$ &$2.9\pm1.1$
&$1.8\pm 0.6$ &$1.5\pm 0.3$\\
theory\tablenote{from Ref.\ \onlinecite{WangKleinKrakauer}} &5.21 &2.08
&2.66\\
exper.\tablenote{$\alpha$-iron from
Ref.\ \onlinecite{GaoJohnstonMurrell}}
&5.406 &2.12 &1.68 &2.33 &1.355 &1.178\\
fcc\\
theory &$6.43\pm 0.1$ &0.0 &$2.6\pm 0.4$ &$5.4\pm 1.0$ &$1.2\pm
0.3$ &$2.6\pm 0.6$\\
theory\tablenote{Low-Spin results from
Ref.\ \onlinecite{WangKleinKrakauer}} &6.37 &0.0 &3.9\\
exper.\tablenote{$\gamma$-iron from
Ref.\ \onlinecite{GaoJohnstonMurrell}} &6.947 &0.0 &1.32 &1.54
&1.22 &0.77\\
\end{tabular}
\end{table}

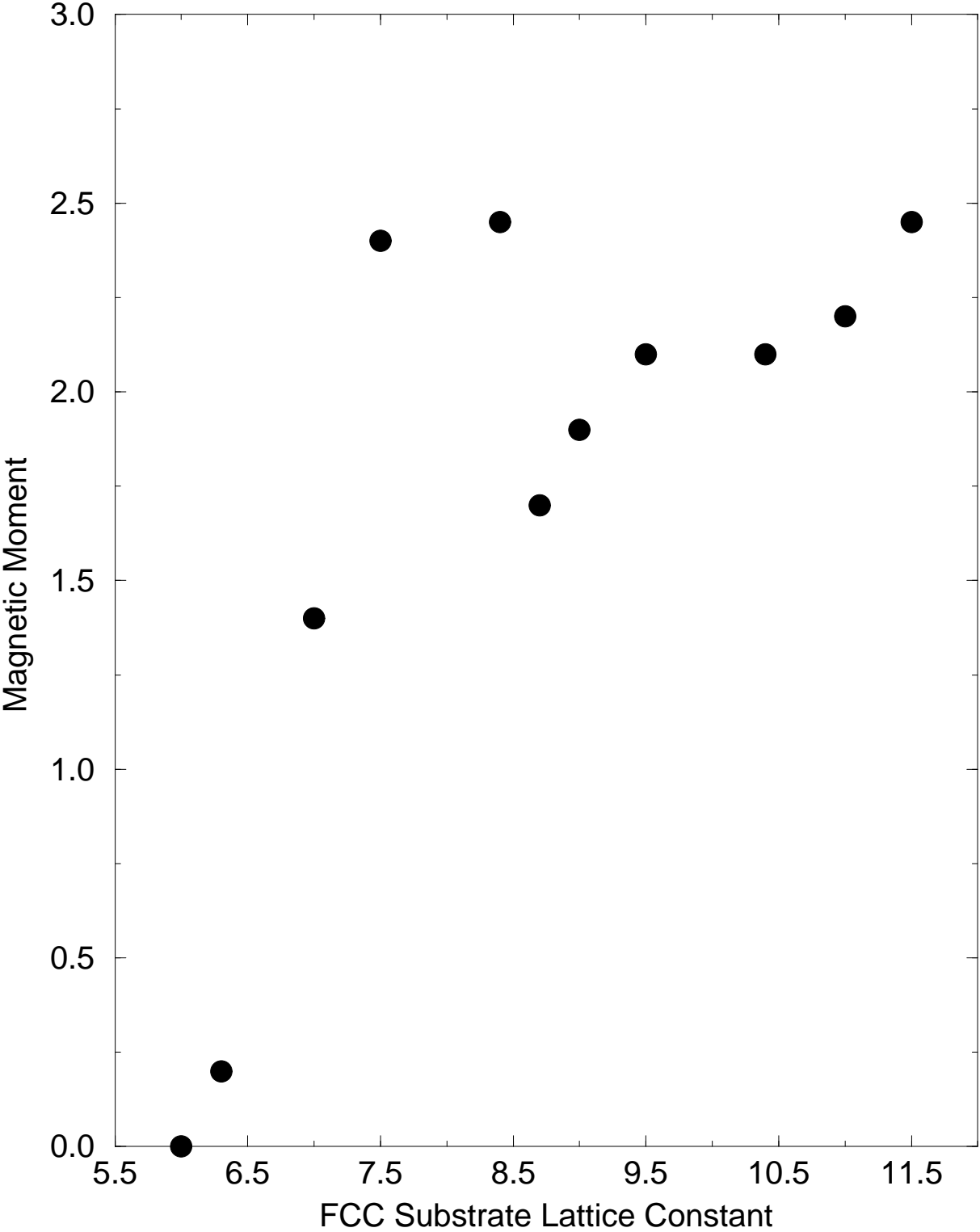
\end{document}